\documentclass{aa}  

\usepackage{graphicx}
\usepackage{txfonts}
\begin{document}

\title{Observational evidence for two-component distributions describing solar magnetic bright points}

\subtitle{}

\author{Gerardine Berrios Saavedra\inst{1},
          Dominik Utz\inst{2,3,4},
          Santiago Vargas Domínguez\inst{5},
          José Iván Campos Rozo\inst{6},
          Sergio Javier González Manrique\inst{7,8,9},
          Peter Gömöry\inst{9},
          Christoph Kuckein\inst{7,8,10},
           Horst Balthasar\inst{10},
            \and
          Peter Zelina\inst{9}
          }

\institute{Universidad Nacional de Colombia, Bogotá, Colombia\\
\email{yberrioss@unal.edu.co}
         \and
         Institute of Physics, Faculty of Science, University of South Bohemia,  \v{C}eské Bud\v{e}jovice, Czech Republic
          \and
Computational Neurosciences, Neuromed Campus, Kepler University Hospital, Linz, Austria\\
\email{dominik.utz@kepleruniklinikum.at}
          \and
          Instituto de Astrofísica de Andalucía IAA-CSIC, Granada, Spain
          \and
Universidad Nacional de Colombia, Observatorio Astronómico Nacional, Bogotá, Colombia\\
\email{svargasd@unal.edu.co}
          \and
University of Graz, Graz, Austria\\
\email{jose.campos-rozo@uni-graz.at}
 \and
Instituto de Astrofísica de Canarias, Tenerife, Spain\\
\email{smanrique@iac.es; ckuckein@iac.es}
 \and
Departamento de Astrofísica, Universidad de La Laguna, La Laguna, Tenerife, Spain
 \and
 Astronomical Institute, Slovak Academy of Sciences, Tatranská Lomnica, Slovakia\\
\email{gomory@astro.sk; pzelina@astro.sk}
 \and
Leibniz -- Institute for Astrophysics Potsdam, Germany\\
\email{hbalthasar@aip.de}
}

   \date{Received --; accepted --}
  
  \abstract
   {High-resolution observations of the solar photosphere reveal the presence of fine structures, in particular the so-called magnetic bright points (MBPs), which are small-scale features associated with strong magnetic field regions of the order of kilogauss (kG). It is especially relevant to study these magnetic elements, which are extensively detected at all moments of the solar cycle, in order to establish their contribution to the behaviour of the solar atmosphere, and ultimately a plausible role within the coronal heating problem.}
   {We aim to characterise the size and velocity distributions of MBPs in the solar photosphere in two different datasets of quiet Sun images acquired with the Solar Optical Telescope SOT/Hinode and the High-resolution Fast Imager HiFI/GREGOR, in the G-band (4308 Å).}
   {In order to detect the MBPs, an automatic segmentation and identification algorithm was used. Next, the identified features were tracked to measure their proper motions. Finally, a statistical analysis of hundreds of MBPs was carried out, generating histograms for areas, diameters, and horizontal velocities.}
   {This work establishes that areas and diameters of MBPs display log-normal distributions that are well fitted by two different components, whereas the velocity vector components follow Gaussians, and the vector magnitude follows a Rayleigh distribution again revealing a two-component composition for all vector elements.}
   {The results can be interpreted as due to the presence of two different populations of MBPs in the solar photosphere, one likely related to stronger network magnetic flux elements and the other one to weaker intranetwork flux elemens. In particular, this work concludes on the effect of the different spatial resolutions of the GREGOR and Hinode telescopes, affecting detections and average values.}
   
   \keywords{Sun: photosphere, evolution. Methods: observational.}
   
\authorrunning{Berrios et al.}
\titlerunning{A two-component distributions describing solar MBPs}
\maketitle

%

\section{Introduction}
\label{sec:1}

High-resolution observations of the solar photosphere reveal a plethora of exceedingly fine structures, mainly corresponding to magnetic bright points (MBPs), which are small-scale features associated with strong magnetic field regions of the order of up to kilogauss \citep[1.5 kG;][]{2003ApJ...587..458N,2007A&A...472..607B,2007A&A...472..911I,2013A&A...554A..65U,2019MNRAS.488L..53K}. MBPs are found all over the photosphere, in both quiet and active regions of the Sun, in particular located in the intergranular lanes in between granular convective cells \citep[e.g.][]{1989SoPh..119..229M}.  Several investigations have found the mean diameter of an MBP to be in the 100 -- 300 km range, its horizontal average velocity to be between 0.2 and 5 \mbox{kms$^{-1,}$} and their average lifetimes to be 2.5 to 10 minutes, as shown by the results of \citet{2004ApJ...609L..91S,2007A&A...472..607B,2010A&A...511A..39U}, and \citet{2018ApJ...856...17L}, among others.\\

According to \citet{2004ApJ...609L..91S}, the presence of MBPs in the solar photosphere was first reported by \citet{1973SoPh...33..281D}. Since then, MBPs have been extensively, but not fully, investigated from the various viewpoints of theory \citep[e.g.][]{1979SoPh...61..363S,1978ApJ...221..368P,1984A&A...139..435D}, observations \citep[e.g.][]{2007ApJ...661.1272B,2009ApJ...700L.145V}, and simulations \citep[][]{2010A&A...509A..76D,2014A&A...568A..13R}. 
Among the better covered aspects of MBPs are their general characteristics as well as their formation, which is believed to be due to the convective collapse theory introduced by \citet{1979SoPh...61..363S} and \citet{1978ApJ...221..368P}. It is also impressively demonstrated in observations presented in a case study by \citet{2008ApJ...677L.145N} and on statistical grounds by \citet{2009A&A...504..583F}. Other authors also tried to identify the formation of MBPs in numerical simulations and compared them with observations (e.g. \citet{2010A&A...509A..76D}). Another hot topic is MBP and magnetohydrodynamics (MHD) waves, as these waves can contribute to the solar atmospheric heating \citep[see, e.g.][]{2009Sci...323.1582J,2009A&A...508..951V,2011ApJ...727...17F}.
Aspects that are similarly important but less extensively covered are the long-time evolution of MBP patterns with the solar cycle and their relation to a possible surface dynamo and/or global dynamo \citep[e.g.][]{2016A&A...585A..39U,2017PASJ...69...98U}. Similarly, the relationship of MBPs and the formation of solar spicules \citep[e.g.][]{2019Sci...366..890S} showed that magnetic flux cancellation between intranetwork fields and network fields represented by MBPs leads to the formation of spicules.

Magnetic bright points  can exist as isolated features, and they can form groups, as reported by \citet{2003ApJ...587..458N} and \cite{1996ApJ...463..365B}. Their dynamics can be seen as highly influenced by the surrounding granulation pattern, and they themselves are being pushed by granules, giving rise to a more or less chaotic movement pattern \citep[e.g.][]{1983SoPh...85..113M,2009Ap&SS.tmp..104S}. \citet{2003ApJ...587..458N} and \cite{1996ApJ...463..365B} also found that two MBPs can come together and form only one (coalescence), or one can divide into two smaller MBPs (fragmentation). In particular, \cite{1996ApJ...463..365B} analysed images in the G band and in the continuum, which were acquired with the Swedish Vacuum Solar Telescope  
\citep[SVTS;][]{Scharmer:85} at the Observatorio Roque de los Muchachos in La Palma, Canary Islands, Spain, finding rapid fragmentation and coalescence in a sample of MBPs under analysis.\\

Magnetic bright points correspond to thin tubes of magnetic flux, as explained by \citet{2001A&A...372L..13S}. These widen in diameter from the photosphere to upper heights. When these magnetic elements are observed in high-resolution filtergrams of the solar photosphere, they look brighter than their surroundings \citep[][]{2005A&A...437L..43Z}. This is due to the higher magnetic field concentration inside the flux tube generating a strong magnetic pressure and, in order to maintain the equilibrium, the gas pressure inside the tube needs to be reduced. This chain of causal effects ultimately leads to lower densities inside the flux tube \citep[see][]{2003ApJ...597L.173S}. The dynamic process to reach the equilibrium is called convective collapse, and more details can be found in the works of \citet{1979SoPh...61..363S} and \cite{1978ApJ...221..368P}. Due to the lower density, the opacity also decreases, giving rise to a higher contribution of emission of radiation from deeper layers and thus increasing the overall brightness of the MBP structure. Added to this, the hot walls of the flux tube can also contribute with an excess of radiation, especially when viewed away from the disc centre \citep[for more details see][]{2001A&A...372L..13S}.\\

The study of these magnetic elements is considered relevant to establishing their plausible contribution to the behaviour of the solar atmosphere, and ultimately to the well-known coronal heating problem \citep[e.g.][among others]{2006SoPh..234...41K,2007AN....328..726E}. As \cite{1993SoPh..143...49C} explained, although small, MBPs could significantly contribute to the energy budget in the corona, as they harbour strong magnetic fields and are spread over the entire surface of the Sun. Theoretically, the movement of the footpoint of an MBP could generate a flow of energy that ascends to the corona and can contribute to its heating \citep[e.g. via the generation of Alfv{\'e}n waves;][]{2009Sci...323.1582J,2013SSRv..175....1M}. Observations of MBPs can also be acquired in other wavelengths. According to \citet{2003ApJ...587..458N}, MBPs and the filigree (chains of bright features located over the intergranular lanes) are observed in lines of the solar spectrum such as H$\alpha$, Ca {\scriptsize II} H, and K; i.e. lines formed in the chromosphere. Furthermore, the \ion{Na}{i} D$_1$ and D$_2$ lines can also be used to identify MBPs, as well as the photospheric \ion{Si}{i} 10827\,\AA\ and \ion{Ca}{i} 10839\,\AA\ lines \citep{2010ApJ...719L.134J, 2019A&A...630A.139K}. \citet{2009CEAB...33...19W} found a preferred location of these structures (that they called magnetic inter-granular structures) at the footpoints of dark H$\alpha$ fibrils, which are known to delineate the network boundaries.\\   
In this work, the analysis of MBPs is done using a time series of images of the solar photosphere acquired with high-resolution, ground-based, and space-borne solar telescopes, that is with the Solar Optical Telescope instrument \citep[SOT;][]{2008SoPh..249..167T} of the Hinode mission \citep[see][]{2007SoPh..243....3K} on the one hand, and the High-resolution Fast Imager \citep[HiFI;][]{2017IAUS..327...20K} placed at GREGOR, an instrument observing in the G band (4308 Å), on the other hand. We focused on the detection of MBPs by means of an automatic segmentation and identification algorithm described in \cite{2014ApJ...796...79U} to further track their evolution and characterise their dynamics and some physical parameters. Section~\ref{sec:2}  describes the datasets and the algorithm used. In Sect.~\ref{sec:3}, we give the results of the analysis. In Sect.~\ref{sec:4}, we discuss the influence of intensity cut-offs and spatial resolution, and in Sect.~\ref{sec:5} we cover the principal discussion. Finally, the main conclusions are presented in Sect.~\ref{sec:6}.\\


\section{Observational data and methodology}
\label{sec:2}

\begin{figure*}
\centering
\includegraphics[width=\textwidth]{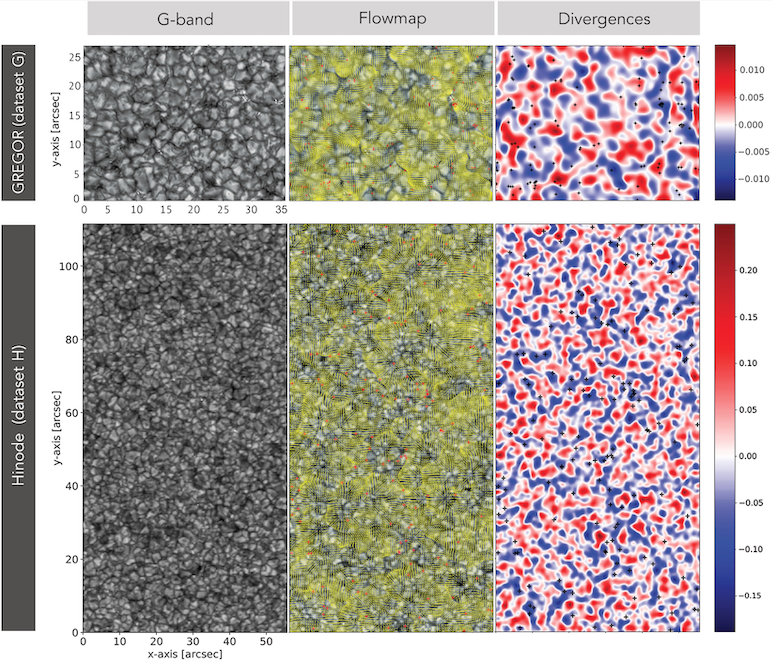}
\caption{G-band images, velocity map, and divergence map (from left to right) for dataset G (top panels) and dataset H (bottom panels). The red and black crosses mark the MBPs identified in the FOV. We note that the spatial scales are different for each dataset.} 
\label{fig:1} 
\end{figure*}

This study makes use of datasets taken on two different days with two different telescopes. The Hinode set (henceforth dataset H) consists of 332 images acquired with the Hinode satellite, specifically with the Solar Optical Telescope \citep[SOT;][]{2008SoPh..249..167T}. These data correspond to solar observations acquired on March 10, 2007, spanning from 7:00 to 9:59 UT. The temporal resolution is 30~s, with a spatial sampling resolution of 0\farcs108 per pixel.
The field-of-view (FOV) is 55\farcs8 $\times$ 111\farcs6 (see bottom-left panel in Fig.~\ref{fig:1}). These observations are not affected by the adverse effects of the atmospheric seeing. On the other hand, the GREGOR telescope \citep{2012AN....333..796S,2020A&A...641A..27K} dataset (henceforth dataset G) consists of 769 images acquired on July 13, 2019 with the HiFI instrument and spanning from 7:38 to 08:48 UT. These data are characterised by presenting a temporal resolution of 5~s, a pixel size of 0\farcs0286 per pixel, and a FOV of 32\farcs4 $\times$ 26\farcs0 (see top-left panel in Fig.~\ref{fig:1}). Despite these observations being affected by atmospheric seeing, corrections made with the adaptive optics system \citep{2012AN....333..863B} considerably improved the image quality. Figure~\ref{fig:1} (left panels) shows the corresponding FOV (in G-band images) for dataset G and dataset H, as labelled.

The first step, before applying the analysis algorithms, is to focus on the data reduction. Dataset H is acquired in zero level of calibration from the freely accessible Hinode database. The calibration is performed with the SolarSoft routine fg$\_$prep implemented in the Interactive Data Language (IDL); the routine executes a general reduction of intrinsic errors including flat field, dark current, and bad pixels. 
Dataset G is calibrated using the data reduction pipeline sTools \citep{2017IAUS..327...20K}, which includes dark-current and flat-field corrections, as well as image selection. Then, speckle interferometry \citep{2008SPIE.7019E..1EW} is used to produce level-2 data, which further corrects aberrations caused by the atmosphere, as explained by \citet{2017IAUS..327...20K}. 

\begin{figure}
\centering
\includegraphics[scale=0.21]{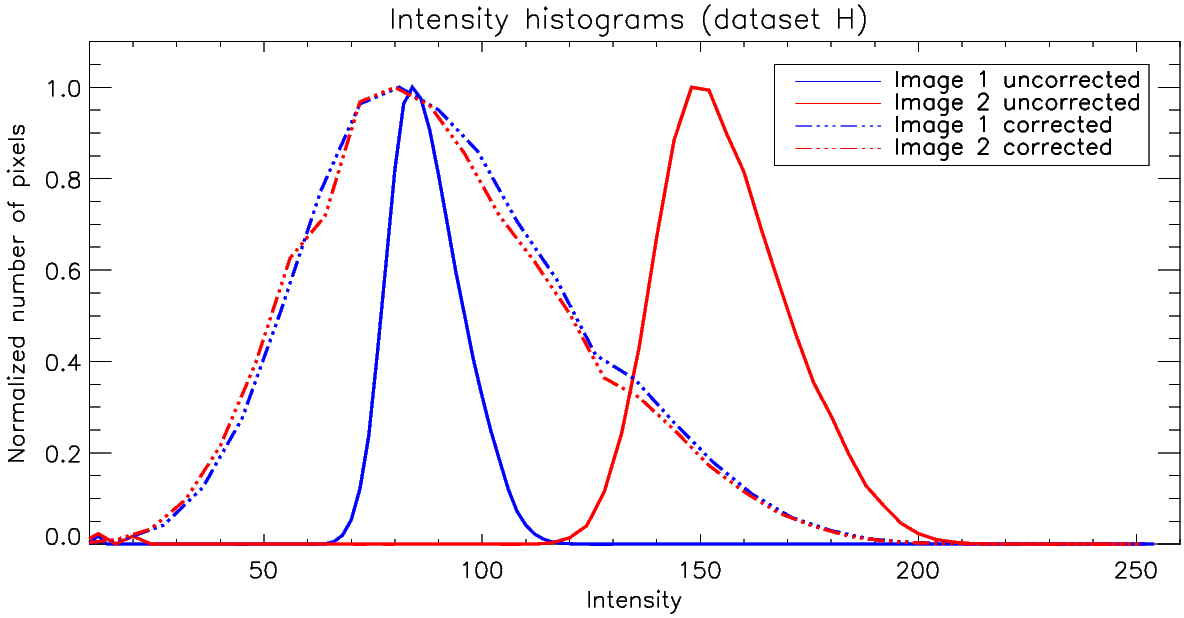}
\caption{Intensity histograms from two images, before (continuous lines) and after (dashed-dotted lines) the intensity re-calibration process is applied.}
\label{fig:2}
\end{figure}

Finally, the two datasets are corrected from intensity disparities. When one evaluates the intensity level of the images organised in a data cube, it is found that each image can vary slightly in the dynamic range of intensities as well as its central intensity. Therefore, it is necessary to perform an intensity re-calibration process that consists of modifying the dynamic range\footnote{Defined here as the difference between the highest percentile of intensity values compared to the lowest percentile.} of the images in such a way that the range remains the same for all images. In the same way, the average intensity of all the images should be set at the same value. In this way, it is guaranteed that all the images have a constant intensity range. Figure~\ref{fig:2} illustrates this process in more detail. The figure depicts two intensity histograms in continuous blue and red lines for images 1 and 2, respectively, before the process of intensity re-calibration; it can be seen that both the dynamic range and the mean intensity are different. The intensity histograms in dashed-dotted style depict the re-calculated histograms after the normalisation process was applied for each image, respectively. It can be proven that both the dynamic range and mean intensity for the two images are statistically similar now, as expected for images of the quiet Sun.

After the pre-processing is completed, the identification of MBPs is made with an automatic algorithm of detection based on the following steps: firstly, each image is segmented in the following way by multiple-thresholding. In the first step, the brightest pixel of the image is detected and initialises the first segment. In the next step, an intensity threshold slightly lower than the intensity level of the brightest pixel is applied,  and the newly found pixels which border the existing segment are added to the existing segment. The remaining pixels (that do not border the existing segment) initialise new segments. After that, the same process is repeated with the succeeding lower intensity threshold. Thus, by going through the whole image intensity range from the brightest to the darkest pixel, the image is broken up into image segments (practically speaking, local intensity maxima initialise segments that grow in size until they reach the boundary of other segments). Subsequently, MBPs in each created segment are detected based on the brightness gradient.

Once the MBPs have been identified (in the case of GREGOR, about 110 MBPs per image and about 170 for Hinode; yielding a total number of detected MBP instances of 87 000 for GREGOR, and 56 000 for Hinode, respectively) their size in pixels is calculated; areas are determined  in \mbox{km$^2$},  and subsequently their diameters calculated in \mbox{km}, assuming that the MBP features have a circular shape, which is considered a good approximation for the purposes of this analysis comparing relative sizes of MBPs when detected by different instruments\footnote{As \citet{2013Ap&SS.348...17F} showed, assuming an elliptical shape would be more precise, as the modal value of the ratio between the major and minor axis of an equisized ellipse would be around 1.5. However, for comparison purposes we follow the tradition of assuming a round or circular shape and stating diameters for equisized circles, as well as the raw results in the form of a size distribution.}. Area and diameter histograms are presented and commented on in detail in Sect.~\ref{sec:3}. Furthermore, the brightness barycentres of each MBP are calculated.

Finally, these tiny elements are tracked in a sequence of images with the purpose of determining their dynamic and evolutionary characteristics. This procedure consists of comparing the intensity barycentre position of one MBP in an image with all the found barycentre MBP positions in the next time instance. If the distance between two positions is less than or equal to one pre-established maximum distance, then the two individual detections of MBPs belong to the the same MBP time sequence.\footnote{Due to this definition,  MBPs that have one temporal realisation are explicitly allowed to have several follow-up MBP realisations; i.e. splitting is allowed and can be detected. The estimation of such splitting and merging events, as well as the occurrence rates of such events will be an interesting topic of a future work.} The maximum distance parameter can be found with the following two considerations: i) the allowed distance must be high enough to allow for the highest possible velocities which can be estimated by, for example, taking 3.5 times the previously established larger MBP velocities of about 4 km/s \citep[e.g. in the work of][one can see a velocity tail starting roughly with 4 km/s extending up to (6 to 8) km/s]{2003ApJ...587..458N}; and ii) by being small enough so that one would not connect a non-involved MBP to the time series. This can be estimated by keeping the maximum distance smaller than an average MBP diameter (about 70 km). Both considerations together represent a search distance of about 70 km or a time cadence of 5 seconds, translating to the maximum detectable velocities of about 14 km/s. By applying these processing steps to all the images and detected MBPs, we obtain the time sequences and evolution of all the MBPs. In order to calculate MBP velocities, barycentre displacements between MBPs identified by the algorithm in subsequent frames are utilised. This leads finally to the estimation of their velocities and their statistical velocity distributions.\\ 
  
\section{Distribution of sizes and velocities of MBPs}
\label{sec:3}

\begin{figure*}
\centering
\includegraphics[width=\textwidth]{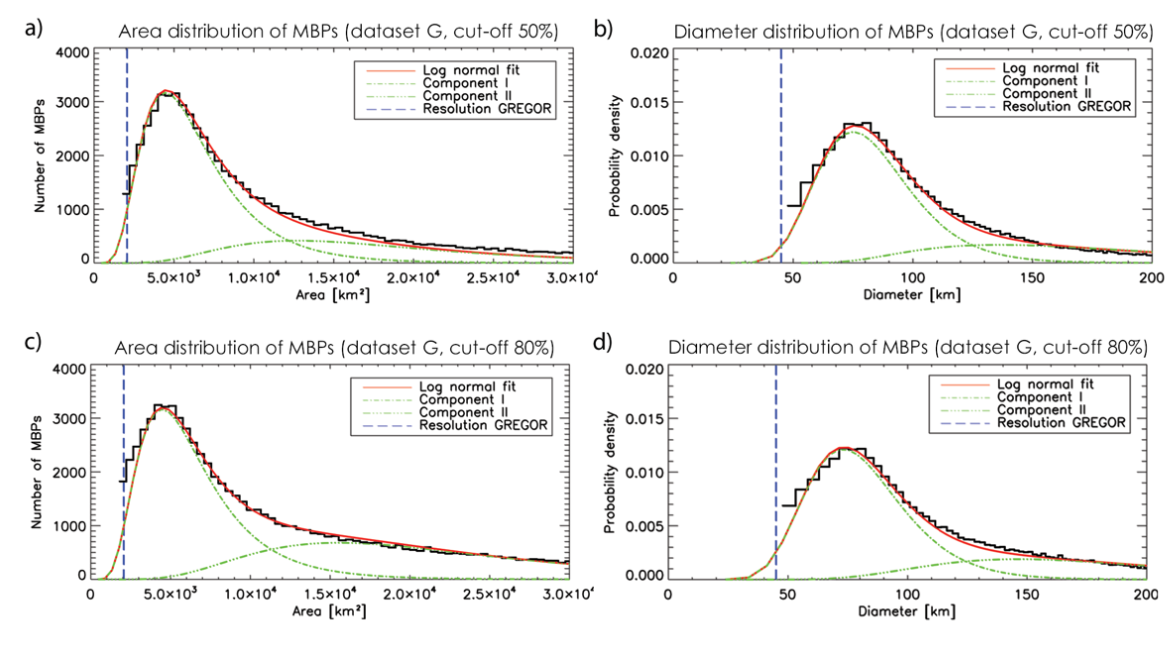}
\caption{Area distributions (panels a and c) and diameter distributions (panels b and d). Both cases are shown for dataset G with cut-offs of 50\% (panels a,b) and 80\% (panels c and d). The plots depict a distribution with a log-normal fit (red curves) and the two components (green curves). The blue vertical line around 2000 $\mbox{km}^2$ (45~\mbox{km}) for area (diameter) indicates the spatial resolution by diffraction of GREGOR.
}
\label{fig:3}
\end{figure*}

\begin{figure}
\centering
\includegraphics[scale=0.11]{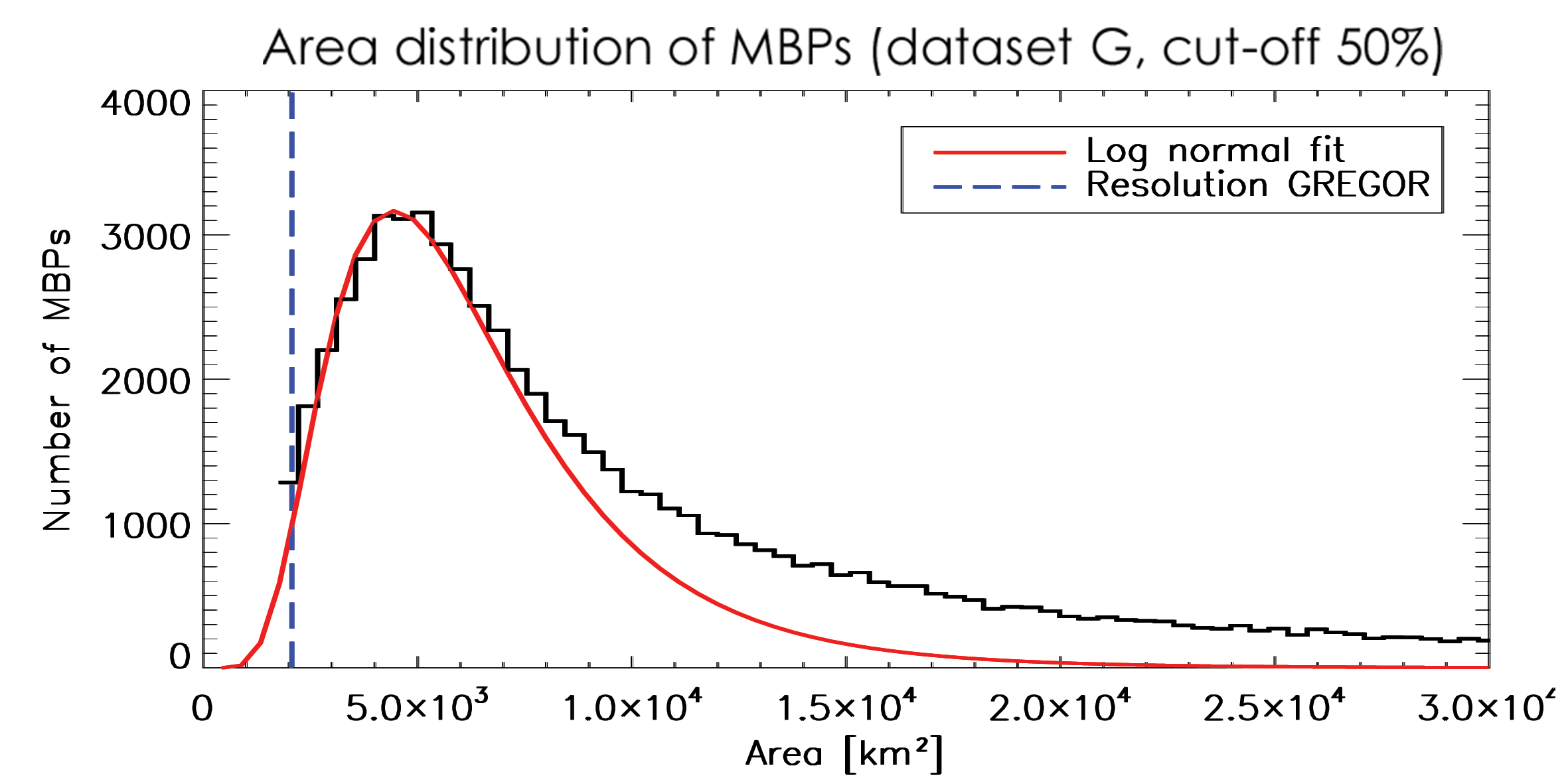}
\caption{Area distribution of MBPs for dataset G (cut-off 50\%) showing a one-component-only fit (in red). Evidently, the tail of the histogram is not well fitted, leading to the idea of two-component fit.}
\label{fig:4}
\end{figure}

In order to characterise the dynamics of the MBPs in the photosphere in quiet Sun regions, this section deals with the analysis of areas, diameters, and velocities, particularly for dataset G as it represents the highest resolution data used in this work. The actual sizes of the MBPs are calculated by applying either a 50\% or 80\% cut-off level on each of the segments identified to be an MBP (for more details, refer to Sect.~\ref{sec:5}). The final histograms of the distribution of areas and diameters are illustrated in Fig.~\ref{fig:3}. Here, we used a binning of the original pixel size of 1 and then re-calculated the corresponding size in square kilometers as well as the diameter in kilometers. As this transformation is a non-linear process (via the square root function), the binning looks quenched to larger diameters in this representation. The red curves in the figure represent the log-normal fits for every distribution, and the curves in green correspond to two underlying different components that constitute the distributions in red and each of these is interpreted as a group of MBPs with well-defined characteristics. As evidenced in Fig.~\ref{fig:4}, the wing is not well represented by a one-component fit, and therefore the need to include a second component to better depict the histogram arises. Moreover, when computing the $X^2$ test values to quantify the goodness of the fit using the observed frequencies (i.e. values obtained in the data) and expected frequencies (i.e. theoretical values set by the fit), similarly to \citet[][]{2010ApJ...722L.188C}, one finds a value of 1.3 for the one-component fit model, which yields a p value of 0.5, meaning that there is only a 50-50 likelihood that the one-component model truly fits the observed data. However, when computing the $X^2$ value for the two-component model (2.6), the picture changes as the p value now reaches  0.8, meaning that there is now an 80 percent likelihood that the model truly fits the measurement values. Practically, a one-component model is just not flexible enough to produce a fit well enough to clearly be able to reproduce the data. The vertical line in blue at 45~\mbox{km} in Fig.~\ref{fig:4} represents the diffraction-limited spatial resolution of the GREGOR telescope at the given wavelength.

The distribution of the areas for 50\% (see panel a in Fig.~\ref{fig:3}) exhibits a mean value of $\mu=5700$ $\cdot\mid\div$ 1.6 $\mbox{km}^2$\footnote{Growth and fragmentation processes that follow multiplicative rules instead of additive rules follow log-normal distributions instead of normal distributions. This means that the additive standard deviation becomes a multiplicative standard deviation. For all the details, see \citet{Eckhard}.} for the first component (I), which corresponds to 73\% of the total distribution, making it a more sizeable group. Due to their sizes, this component likely corresponds to newborn MBPs or MBPs in formation, with an area range of 2200 -- 15000 $\mbox{km}^2$. As for the second component (II), the mean value is $\mu = 16000$ $\cdot\mid\div$ 1.6 $\mbox{km}^2$, which corresponds to 27\% of the total; this component also represents a family of MBPs; however, these are mature ones, which have bigger areas in the 6300 -- 40000 $\mbox{km}^2$ range but are decreased in quantity. 
Table~\ref{table:1} lists the mean values of the distributions together with the factor that multiplies and divides it. The factor (e.g. the value 1.6) is proper to this type of curve, representing the value that multiplies and divides the mean value and indicating the shape of the distribution \citep[log-normal; for more details, see][]{Eckhard}. These results are in agreement with previous studies \citep[e.g.][]{2009A&A...498..289U,2010ApJ...722L.188C}, which established values between 7000 and 70000 $\mbox{km}^2$ for the area of MBPs. Although our obtained mean value of 5700 $\mbox{km}^2$ is smaller than some of the previous calculations, it can be understood as evidence for the higher resolution of the GREGOR telescope affecting the detection procedure. For instance, \cite{2010ApJ...722L.188C} found a peak in the area distribution at 45000 km$^2$ using both G-band observations from the 76 cm Dunn Solar Telescope (DST) with a diffraction limit at 10000 km$^2$ and numerical simulations using the MURaM code \citep{2005A&A...429..335V}.

Similarly, the distribution of diameters (see panel b in Fig.~\ref{fig:3}) has a mean value of $\mu = 80$ $\cdot\mid\div$ 1.3 $\mbox{km}$ for 70\% of the total distribution in the first component. Compared to that, the second component has a mean diameter of $\mu = 130$ $\cdot\mid\div$ 1.6 $\mbox{km}$ for 30\% of the distribution (see Table~\ref{table:1}). Once more, each component (dotted green lines in Fig.~\ref{fig:3}) of this distribution depicts a population of small MBPs or in MBPs in the process of formation or disintegration. The first of these populations, with a greater number of cases, remains in a diameter range of  50-140~\mbox{km}. Moreover, the second population of MBPs, which is characterised as having fewer elements, exhibits diameters spanning a range from 50-330~\mbox{km}.\\

\begin{figure*}
\centering
\includegraphics[width=\textwidth]{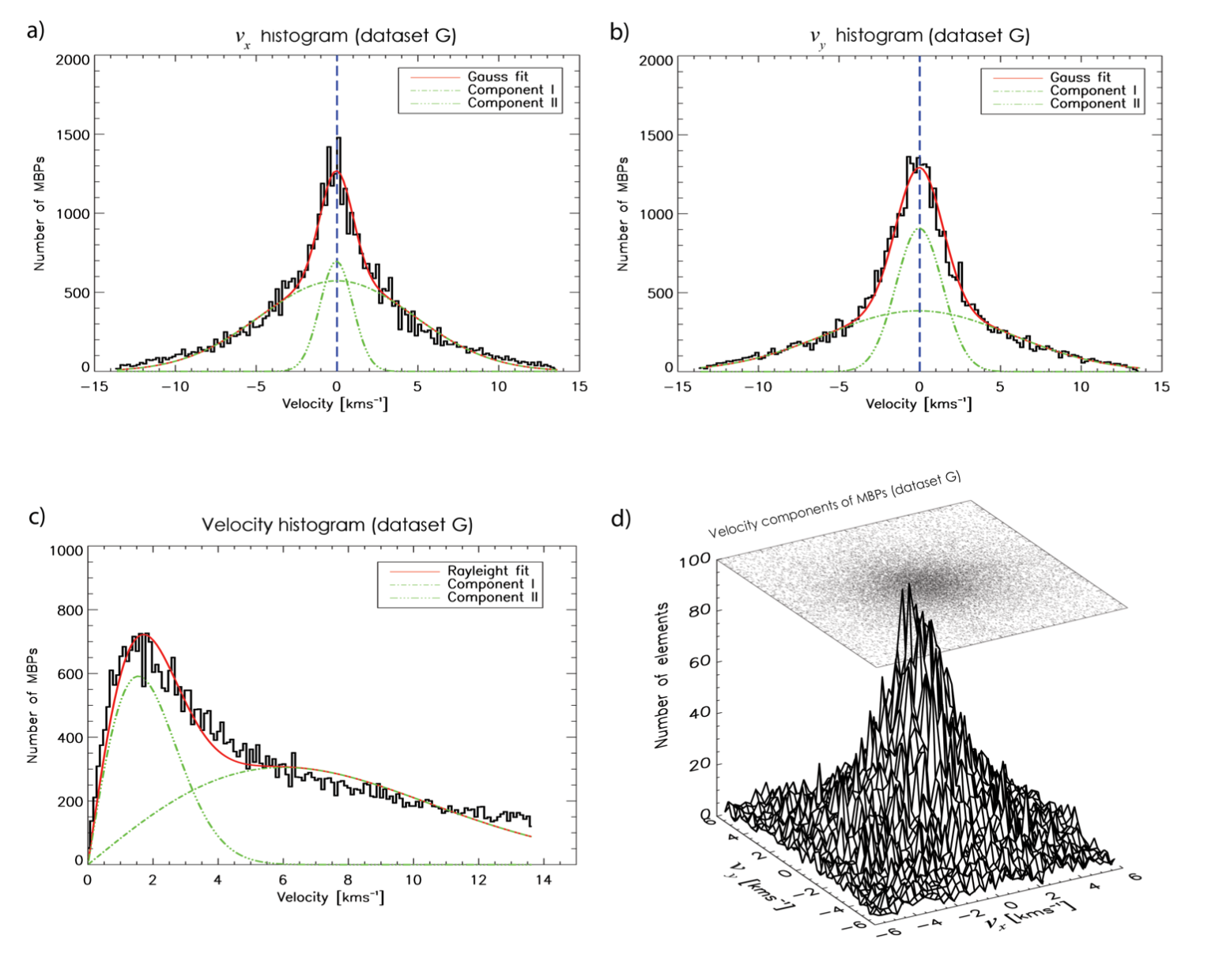}
\caption{Velocity histograms of MBPs for the GREGOR set. Top: $v_x$ and $v_y$ velocity components with the normal fit curve (red) and the two components of the distribution (green). Bottom left: Velocity with the Rayleigh fit in red and the two components in green. Bottom right: Surface and scatter plot of the $v_x$ and $v_y$ components.}
\label{fig:5}
\end{figure*}

Figure~\ref{fig:5} shows the velocity distribution for the same dataset in their velocity components $v_x$ and $v_y,$ together with the surface and scatter plots. Here, and in the following, positive values for $v_x$ represent movement to the right, while a positive value for $v_y$ means a movement to the top of the image (images oriented with solar north to the top of the image). The analysed velocities were calculated in a frame-by-frame or MBP-by-MBP instance, without considering any temporal averaging (except of the unavoidable averaging by the given cadence of the data). The velocity histogram, shown in panel c in the figure, is fitted by a Rayleigh function (red) that can be split into two components (green). The mean value for the first component (I) of the velocity (see panel c in Fig.~\ref{fig:5}), which represents the 67\% of the total distribution, is $\mu = 7.5$ $\mbox{kms}^{-1}$. For the second component (II), the mean value is $\mu = 1.9$ $\mbox{kms}^{-1}$ with 33\% of the total. Each component plausibly indicates a different population of MBPs (discussed in Sect.~\ref{sec:4}), the first of which exhibits a velocity in the 0-13~$\mbox{kms}^{-1}$ range and is more numerous. Accordingly, the second component depicts a population of MBPs with a 0-5~$\mbox{kms}^{-1}$ velocity range with fewer elements. Unlike values for areas and diameters, velocity components $v_x$ and $v_y$ are well fitted by Gaussian distributions. Panels a and b in Fig.~\ref{fig:5} show the frequency histograms of the velocity in $x$ ($v_x$) and in $y$ ($v_x$) with, again, evidence for a behaviour described by two components (component (I) and component (II), as labelled for the green curves). A blue vertical line was located at the zero velocity value for clarity. In the case of the $v_x$ , the mean value is found to be $\mu = 0.009 \pm 4.8$ \mbox{kms}$^{-1}$ for component (I), with 80\% of the total distribution. This component represents a population of MBPs with a range from -14-14 \mbox{kms}$^{-1}$ and is characterised by a higher number of elements. On the other hand, component (II) has a mean value of $\mu = -0.14 \pm 1$ \mbox{kms$^{-1}$} , with 20\% of the total (as listed in Table~\ref{table:2}), and corresponds to a population of MBPs in a -3-3 \mbox{kms}$^{-1}$ velocity range: moreover, it is characterised by a lower number of elements. Analogously, the distribution of $v_y$ has a mean value for component (I) of $\mu = -0.08 \pm 5.7$ \mbox{kms}$^{-1}$ with a total contribution of 63\%, representing a population of MBPs featuring velocities within a -17-17 \mbox{kms}$^{-1}$ range. This component features a higher number of elements, whereas component (II) has $\mu = -0.03 \pm 1.4$ \mbox{kms}$^{-1}$ with a 37\% contribution to the total distribution, representing a population of MBPs featuring velocities in the -4-4 \mbox{kms}$^{-1}$ range. This second component is characterised by a lower number of elements. Both distributions, $v_x$ and $v_y$, show similar behaviour, as is be evidenced in Fig.~\ref{fig:5}. In the same figure, it can be noticed that the underlying population of MBPs for component (II) of velocity distributions corresponds to component (I) for area and diameter distributions. This behaviour can be interpreted as MBPs of smaller sizes belonging to a population that exhibits a more extended velocity distribution and thus larger velocities on average. Contrarily, larger MBPs are characterised as having lower velocities, i.e. component (I) velocity distributions correspond to component (II) distributions for area and diameter. Panel d in the figure serves to display a 3D plot combining the information from $v_x$ and $v_y$ values, and a scatter plot that better evidences the central concentration of points. 

\begin{table}
\caption{List of areas and mean diameters for the different datasets (G and H), cut-offs (50 and 80\%), and components (I and II).}
\centering
\begin{tabular}{cccccc}\hline
Dataset & \% & Component & Mean area & Mean diameter\\
    &    &           &    \mbox{(km$^2)$} & \mbox{(km)}\\\hline
G & 50 & I & 5700 $\cdot\mid\div$ 1.6 & 80 $\cdot\mid\div$ 1.3\\ 
G & 50 & II & 16000 $\cdot\mid\div$ 1.6 & 130 $\cdot\mid\div$ 1.6\\
G & 80 & I & 5700 $\cdot\mid\div$ 1.6 & 80 $\cdot\mid\div$ 1.3\\
G & 80 & II & 20000 $\cdot\mid\div$ 1.6 & 150 $\cdot\mid\div$ 1.5\\\hline
H & 50 & I & 230000 $\cdot\mid\div$ 1.9 & 510 $\cdot\mid\div$ 1.7\\ 
H & 50 & II & 494000 $\cdot\mid\div$ 1.2 & 780 $\cdot\mid\div$ 1.3\\
H & 80 & I & 37000 $\cdot\mid\div$ 1.7 &  260 $\cdot\mid\div$ 1.5\\
H & 80 & II & 115000 $\cdot\mid\div$ 1.2 & 380 $\cdot\mid\div$ 1.2 \\\hline
\end{tabular}
\label{table:1}
\end{table}
   
\begin{table*}
\caption{Mean velocity values (in \mbox{kms$^{-1}$}) for the two datasets (G and H) and components (I and II). The fit parameters (st.dev.) $\sigma$ are indicated.}
\centering
\begin{tabular}{cccccccc}\hline
Dataset & Component & Mean $v_x$ & $\sigma_{v_x}$ & Mean $v_y$ & $\sigma_{v_y}$ & Mean effective velocity & $\sigma_{v_R}$\\\hline
G & I & 0.009 & 4.8 & -0.08 & 5.7 & 7.5 & 6.1\\ 
G & II & -0.14 & 1 & -0.03 & 1.4 & 1.9 & 1.5\\\hline
H & I &  -0.063 & 1.7 & 0.024 & 1.5 & 2.3 & 2.8\\
H & II & -0.00007 & 0.7 & -0.024 & 0.6 & 1.1 & 1.4\\\hline
\end{tabular}%
\label{table:2}
\end{table*}
%
%

\section{Influence of intensity cut-off and spatial resolution: GREGOR versus Hinode}
\label{sec:4}

\begin{figure}
\centering
\includegraphics[scale=0.16]{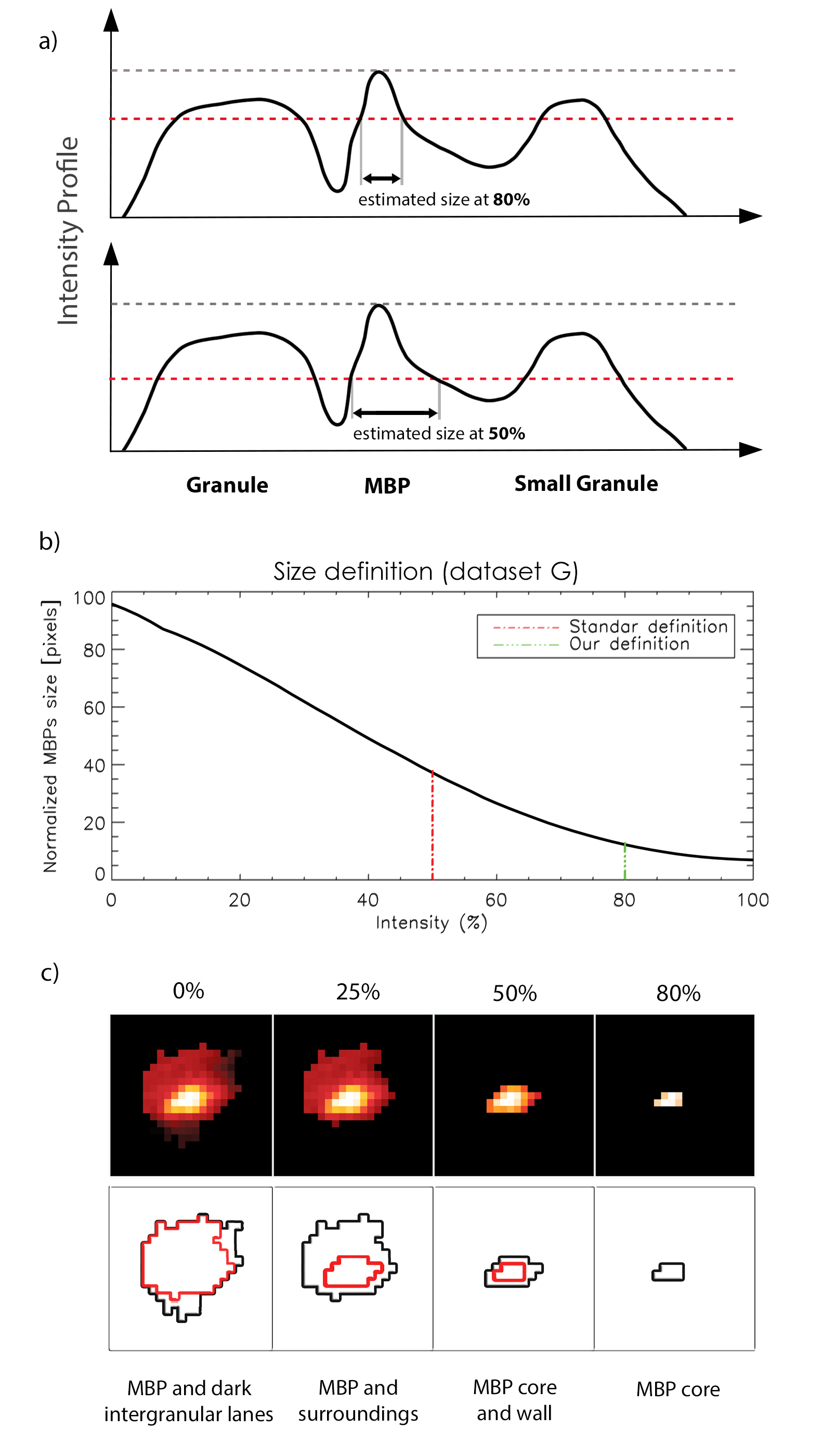}
\caption{Intensity characterisation of MBPs. Panel a: Schematic side view cut of the solar surface in intensity with two granules and an MBP. The dashed horizontal line illustrates the cutting level and the following size of a segmented MBP. Panel b: Criteria for setting the area of an MBP. The concept that underlies this criterion is based on finding a range of independence between the size of the MBP and the intensity cut-off percentage. Panel c: Intensity threshold (for values of 0, 25, 50, and 80\%) applied over a segmented sample image and the effect on the identification of features, as labelled. The lower row represents contours in black for each frame in the upper row, and in red for every corresponding frame on the right, in order to compare the progressive area enlargement as the intensity threshold reduces from 80 to 0\%. For an exhaustive discussion, refer to the main text.}
\label{fig:6}
\end{figure}

In the results described in Sect.~\ref{sec:3} for dataset G,  a cut-off at 50\% is used, that is to say, the commonly employed value by previous similar studies \citep[see e.g.][]{1995ApJ...454..531B}. However, it is necessary to evaluate whether that definition, which is commonly used, yields the most accurate results. In order to do so, the definition that is established is based on finding the existing relationship between the size of MBPs and the percentage of intensity cut-off.
Figure~\ref{fig:6} represents the main scheme behind the characterisation of MBPs according to the intensity profile, displaying the enlargement in identified areas with the reduction of the intensity cut-off. The plot in panel b of Fig.~\ref{fig:6} evidences that the more horizontal the curve is, the more independent the MBP size will be of the cut-off percentage parameter. Panel c shows the evolution of identified areas with the intensity threshold in a sample image, for cut-off values of 0, 25, 50, and 80\%, respectively. Black contours in the lower row highlight the corresponding areas, while red contours are overplotted to compare the area enlargement from one cut-off value to a lower one. They therefore affect the recognised features, i.e. the MBP core, MBP core and wall, the whole MBP and surroundings, and the MBP and some dark intergranular lanes. In panel b of the figure, the stability zone is evidenced in the 60-100\% range. A median of 80\% is the final representative value selected as the cut-off percentage \citep[see also ][who used a similar approach to `shrink' detected granules to correct sizes]{2001SoPh..201...13B}. The description of the results using this new cut-off value is performed to compare cut-off values at both 50\% and 80\%.
The examination of areas for this cut-off result in a two-component distribution with log-normal fits, as shown in panel c of Fig.~\ref{fig:3}, with the first component having a mean value of $\mu = 5700$ $\cdot\mid\div$ 1.6 $\mbox{km}^2$ and representing 57\% of the total distribution. Component (I) corresponds to a large family of newborn MBPs under formation, whose areas cover a range from 2200 to 15000~$\mbox{km}^2$. The other component is characterised by a mean value of $\mu = 20000$ $\cdot\mid\div$ 1.6 $\mbox{km}^2$ and depicts 43\% of the total distribution; this second component corresponds to a family of mature MBPs with areas ranging from 7800 to 52000 $\mbox{km}^2$ but with a smaller number of elements. The previous results are within the range of other studies, such as those described in \citet[][]{2004A&A...422L..63W,2009A&A...498..289U,2010ApJ...725L.101A,2018ApJ...856...17L}, among others, with established values between 7000 and 70000 $\mbox{km}^2$.

As for the distribution of diameters, shown in panel d in Fig.~\ref{fig:3}, the mean value corresponds to $\mu = 80$ $\cdot\mid\div$ 1.3 $\mbox{km}$, which is 70\% of the total, for the first component; this corresponds to a group of MBPs with a range of diameters between 50 and 140 \mbox{km} and with a large number of elements. The second component has a mean value of $\mu = 150$ $\cdot\mid\div$ 1.5 $\mbox{km}$, representing 30\% of the distribution; this corresponds to a group of MBPs with diameters from 70 to 340 \mbox{km} and a lower number of elements. Table~\ref{table:1} summarises the information for mean areas and diameters. In panels c and d in Fig.~\ref{fig:3}, the vertical blue line around 2000~$\mbox{km}^2$ (45 \mbox{km}) represents the spatial resolution limited by the diffraction of GREGOR for areas and diameters, respectively. 
The results for area and diameter for dataset G do not change significantly when raising the cut-off value from 50\% to 80\%, as shown in panels a versus c and b versus d of Fig.~\ref{fig:3}. This indicates the superb quality of the GREGOR telescope and its HiFi instrument, as well as the proper application of the identification and analysis algorithms, as the results are practically independent of the chosen threshold parameter.

\begin{figure*}
\centering
\includegraphics[width=\textwidth]{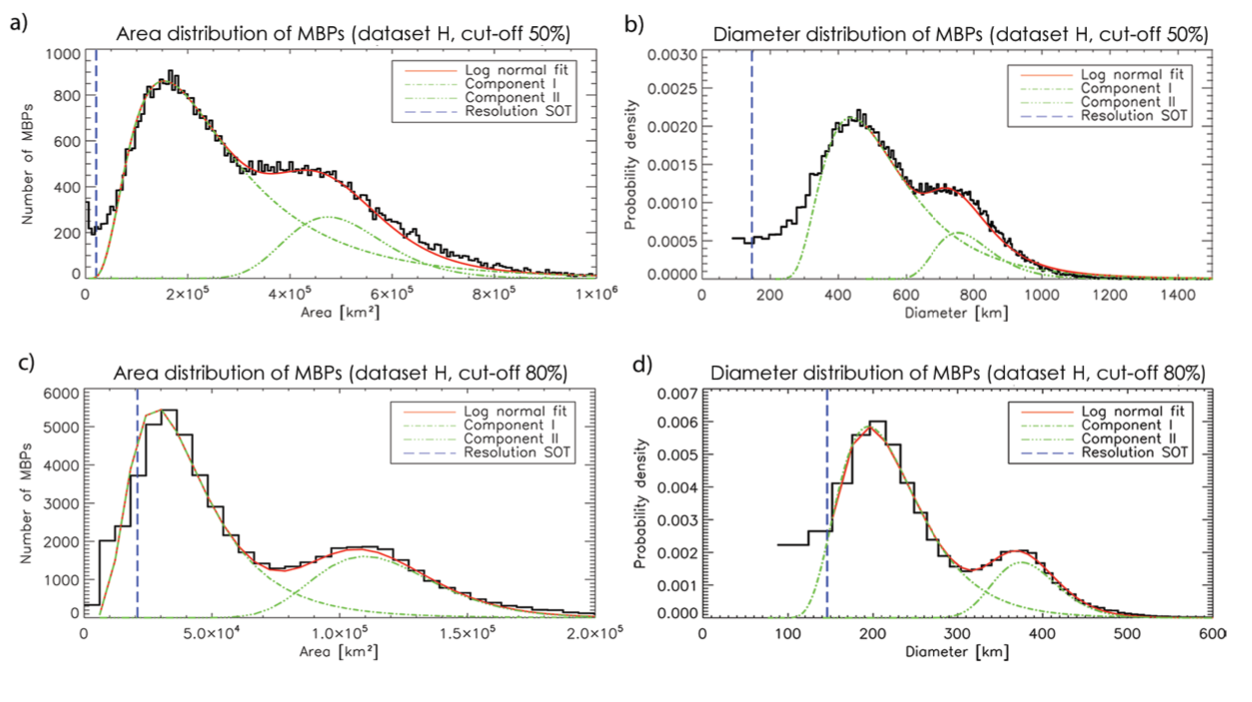}
\caption{Area distributions (panels a and c) and diameter distributions (panels b and d). Both cases are shown for dataset H with a cut-off of 50\% (panels a and b) and 80\% (panels c and d). The plots depict a bi-modal distribution with a log-normal fit (red curves) and the two components (green curves). The vertical blue line at 146~\mbox{km} (around 21000 $\mbox{km}^2$) indicates the spatial diffraction limit of Hinode.}
\label{fig:7}
\end{figure*}

Nevertheless, for the case of dataset H, in which the spatsurpassial resolution is not as high, the results are strongly influenced by the threshold, as plotted in Fig.~\ref{fig:7}. Similarly to dataset G, these distributions also exhibit a bi-modal contribution with two protruding peaks (components (I) and (II)). The fit parameters of the log normal fits are shown in Table~\ref{table:1}. For the case of a cut-off at 80\%, shown in panel c of Fig.~\ref{fig:7}, the mean area of the first component is $\mu=37000$ $\cdot\mid\div$ 1.7 $\mbox{km}^2$ and represents 71\% of the total distribution. On the other hand, the mean value of the second component is $\mu=115000$ $\cdot\mid\div$ 1.2 $\mbox{km}^2$ and corresponds to 29\% of the total distribution. Values to the left of the vertical blue line are not reliable as they lie below the spatial resolution by diffraction for Hinode (i.e. 21000 $\mbox{km}^2$). Considering that the second component of the distribution surpasses the values reported in other studies, e.g. areas from 7000 to 70000 $\mbox{km}^2$ in \citet{2009A&A...498..289U,2010ApJ...722L.188C,2010ApJ...725L.101A,2014RAA....14..741Y}, among others, it is compelling to discuss the meaning of it. A plausible explanation relies on representing two different populations of elements, one with areas in the 22000-63000 $\mbox{km}^2$ range that constitutes a family of MBPs (with greater number of elements), and the second with values from 96000-138000 $\mbox{km}^2$ , which is likely to correspond to a family of small granules in the process of formation or segments of granules, with a lower number of elements, considering the granules have values of area  in the 105000-2600000 $\mbox{km}^2$ range \citep{1980SoPh...65..207K}.
 
For the case of diameters computed with the same cut-off value (80\%), the mean value of the first component of the bimodal distribution, plotted in panel d) of Fig.~\ref{fig:7}, is $\mu=260$ $\cdot\mid\div$ 1.5 \mbox{km} and corresponds to the 84\% of the total. It also has a greater of number elements. Diameters are in the 170-390 \mbox{km range}, in agreement with the values reported by other studies \citep{2009A&A...498..289U}, which are between 100 and 300 \mbox{km}. On the other hand, the mean value of component (II) is $\mu=380$ $\cdot\mid\div$ 1.2 \mbox{km}, which corresponds to a population of forming granules (or fragments) with values from 310 to 450 \mbox{km} and lower number of elements. This result is comparable to the 370-1800 \mbox{km} range found by \citet{1980SoPh...65..207K}.

To verify the precision in the definition of size described above, the distribution of the areas and diameters is calculated using the common definition (intensity cut-off at 50\%). Panels a and b in Fig.~\ref{fig:7} show the two resulting bi-modal distributions that are fitted with log-normal curves (red), together with the two components that constitute it (green). Table~\ref{table:1} lists the fit parameters of such distributions. The mean value for the area of the first component is $\mu=230000$ $\cdot\mid\div$ 1.9 $\mbox{km}^2$ and represents 80\% of the total distribution. For the second component, the value is $\mu=490000$ $\cdot\mid\div$ 1.2 $\mbox{km}^2$, which corresponds to 20\% of the total. Surprisingly,  component (I) of this distribution, with higher number of elements, now shows areas in the 120000 - 400000 $\mbox{km}^2$ range. Therefore, the threshold level in the case of Hinode changed the estimated size of the features quite
dramatically. One has to keep in mind here that the same segments were identified in both cases and only the 'estimation' of the true size was changed by the parameter. A possible explanation for the effect of larger sizes is shown in Fig.~\ref{fig:6}. Panel a shows a schematic representation of the solar surface in a side-cut view. In between two granules, we can find an MBP situated in the intergranular lane. The segmentation algorithm will grow all segments starting from the brightest pixel until the growing segments cease growing when they encounter a neighbouring segment at the darkest intergranular pixels. Thus, an MBP segment can contain quite a large extent of the intergranular area and needs, after segmentation, to be cut down to the brightest part (corresponding to the true size of the MBP). If the cut-off, however, is chosen too low, the MBP size can be artificially extended by intergranular dark lane pixels. One can see that varying the cut level by about 20\% of the brightest intensity does not lead to a significant change of size, as shown in panel b of Fig.~\ref{fig:6}. This means that the brightest pixels of an MBP (the core structure) have homogenous brightness distributions, and thus they are under equivalent physical conditions. Taking the threshold to lower brightness levels, the structure increases in size as more pixels surrounding the brightest core of the structure are added. It is preferable to choose a sufficiently high threshold. It should also come from a part of the curve where it is nearly horizontal, as this corresponds to greater independence of the estimated size from the chosen threshold parameter.
Thus, the first component is also truly an MBP
size distribution at the  50\% cut-off level; however, sizes are estimated too large by a cut level that is too deep. The second component found in the Hinode data (Fig.~\ref{fig:7}) is interpreted as corresponding to a family of granules with area in the 400000-600000 $\mbox{km}^2$ range, but with fewer elements. These results are in agreement with the range found by \citet{1980SoPh...65..207K}, of 105000 to 2600000 $\mbox{km}^2$. By visually inspecting some of the images, we confirm the second distribution is in agreement with false detection of small granules misinterpreted as MBPs. The analysis proves that the 80\% cut-off value manifests greater detection accuracy i.e. 22000 to 62000 $\mbox{km}^2$, whereas results based on the common definition provide very large sizes for the MBPs, of the order of 122000 to 500000 $\mbox{km}^2$. Table~\ref{table:1} compiles the fitted values. Mean value of the first component of this bimodal distribution is $\mu=510$ $\cdot\mid\div$ 1.7~\mbox{km} and represents 81\% of the totality of the same, and for the second distribution it is $\mu=780$ $\cdot\mid\div$ 1.3~\mbox{km} with 19\%.

We can conclude that the resolution of Hinode is not sufficient to detect the bi-component composition of the size distribution of MBPs. Due to the lower resolution, there is a certain level of misidentified forming granules detected in the Hinode dataset constituting the second component in the case of MBP size distribution for dataset H.

\begin{figure*}
\centering
\includegraphics[width=\textwidth]{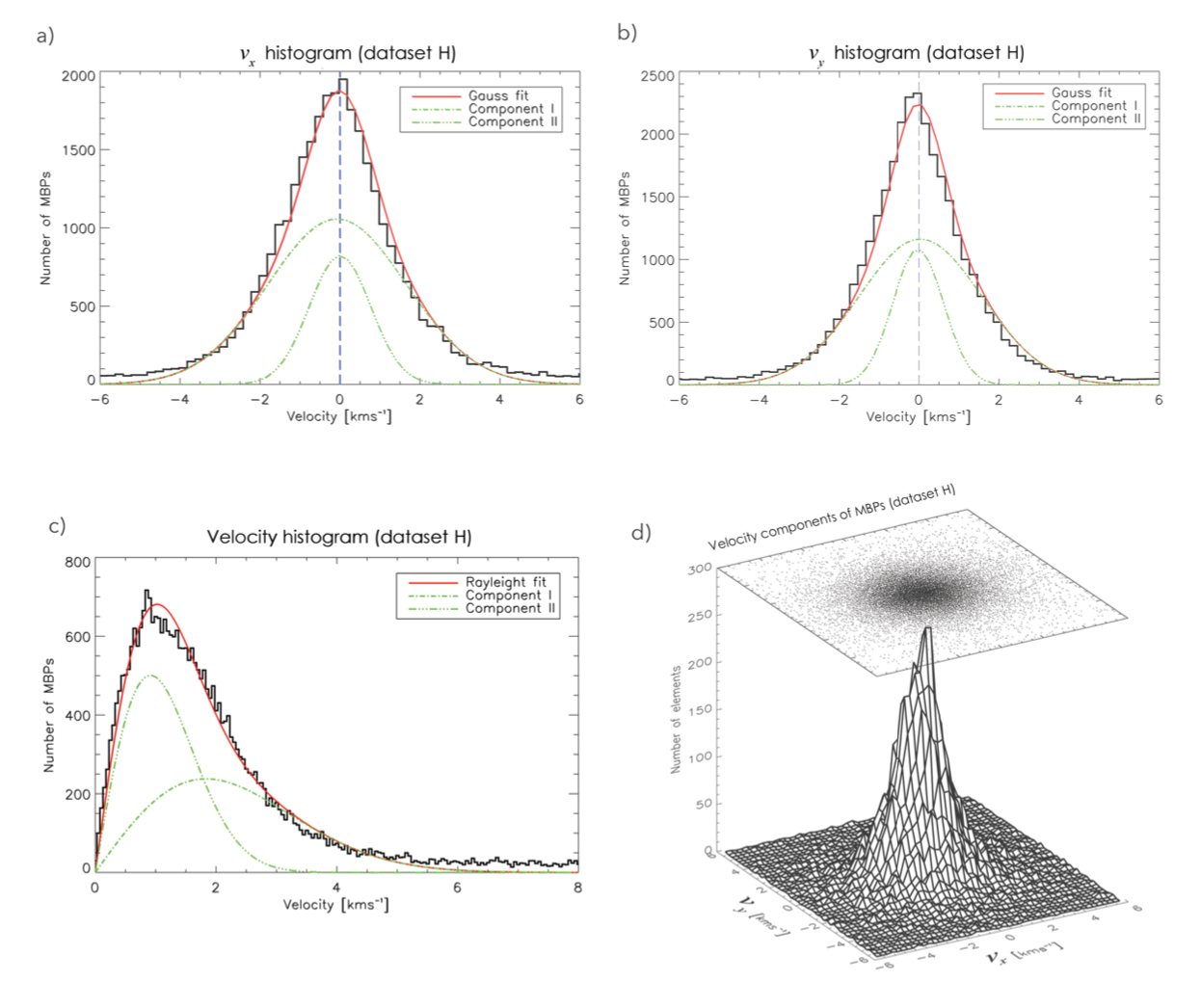}
\caption{Velocity histograms of MBPs for the Hinode set. Top: $v_x$ and $v_y$ velocity components with the normal fit curve (red) and the two components of the distribution (green). Bottom left: Velocity with Rayleigh fit in red and the two components in green. Bottom right: Surface and scatter plot of the $v_x$ and $v_y$ components.}
\label{fig:8}
\end{figure*}

Figure~\ref{fig:8} shows the results for the computed velocities. Panels  a and b in the figure depict the frequency histograms for the components $x$ and $y$ of the velocity, i.e. $V_x$ and $V_y$, respectively, with the fitted curves in red (normal distribution) and the two components represented by green dotted curves. The fit parameters $\mu$ and $\sigma$ are summarised in Table \ref{table:2}. Panel c in the figure displays the effective velocity fitted by Rayleigh distribution (red) and the two components (green). The  mean velocity value for component (I), which represents 49\% of the total distribution, is $\mu = 2.3$ \mbox{kms}$^{-1}$ and is characterised as displaying a velocity range from 0 to 6~$\mbox{kms}^{-1}$ and having a slightly smaller number of elements. Component (II) has a mean value $\mu = 1.1$ \mbox{kms}$^{-1}$ and represents 51\% of the total distribution, indicating a population of MBPs with a velocity range from 0 to 3~$\mbox{kms}^{-1}$ and a slightly higher number number of elements. 

We expressed the velocity components in two orthogonal directions (x, y; horizontal and vertical directions), which correspond to  $v_x$ and $v_y$, respectively, displaying Gaussian distributions as it is expected in the case of a non-preferential direction random walk process. A different outcome would only be expected if there were reasons for a large-scale flow field with preferred directions as in the moat flows of sunspots. Panels a and b in Fig.~\ref{fig:8} show the frequency histograms of $v_x$ and $v_y$, respectively, fitted to normal curves (red), and the corresponding two components (green). A vertical line (blue) is located at the value of velocity equal to 0 \mbox{kms$^{-1}$} for clarity. The mean value for $v_x$ is $\mu = -0.063 \pm 1.7$ kms$^{-1}$ for component (I) (76\% of the total) and represents a population of MBPs in a -5-5 $\mbox{kms}^{-1}$ range, with greater number of elements. For component (II), the mean value is $\mu = -0.00007 \pm 0.7$ kms$^{-1}$, hence 24\% of the total distribution, representing a population of MBPs in a range from -2.2 to 2.2 $\mbox{kms}^{-1}$ with fewer elements. For $v_y$, the mean value is $\mu = 0.024 \pm 1.5$ \mbox{kms}$^{-1}$ for component (I) (73\% of the total) and $\mu = -0.024 \pm 0.6$ \mbox{kms}$^{-1}$ for component (II) (27\% of the total distribution). In panels a and b in Fig.~\ref{fig:8}, both $v_x$ and $v_y$, exhibit essentially the same distribution for the two components. Component (I) represents a population of MBPs with $v_y$ from -4.5 to 4.5~\mbox{kms}$^{-1}$ and a slightly higher number of elements, whereas component (II) indicates a population of MBPs with $v_y$ from -1.9 to 1.9 $\mbox{kms}^{-1}$ and fewer elements.\\ 

\section{Discussion}
\label{sec:5}

It is well known that the magnetic field inhibits convection, and therefore plasma velocities are expected to reduce as the magnetic field intensifies. Thus, it is highly likely that this physical mechanism underlies the two-component behaviour. Some previous investigations have dealt with velocity distributions of MBPs in different environments. \citet{2017ApJS..229...10J} showed the variation in diffusion of MBPs across a network cell (with regions of flux emergence), whereas \citet{2014A&A...566A..99K} demonstrated the velocity distribution in three regions of varying background flux and show that MBPs in regions of higher flux have lower transverse velocities. The authors suggest that this is due to convection being disrupted in regions of higher flux, resulting in more regions of stagnation points for the MBPs reducing velocities. The regions of disrupted convection would relate to the small granules that are mentioned in this work. Convective cells are smaller in regions of higher flux, which in turn creates more regions where there are several intersecting intergranular lanes, forming the stagnation points where MBPs become trapped, thus resulting in a lower peak-velocity distribution. We expect that the differences observed in our analysis are due to the location where the MBPs are found in the quiet-Sun regions (i.e. whether they can be classed as network or internetwork MBPs), but this requires further analysis of velocity flow maps, with the precise testing of parameters and comparison with magnetograms, for a definitive statement to be made. This will be the focus of a subsequent study.

The first group of MBPs, exhibiting lower velocities, are likely to harbour strong magnetic fields that impede the rapid movement of this population. On the other hand, the second group, with greater velocities, would correspond to weaker magnetic fields letting them move faster while they are embedded in the photospheric plasma flow. There is also a recent finding suggesting a bi-component composition magnetic field strength distribution of MBPs, as evidenced by \citet[][]{2019MNRAS.488L..53K}. Figure~\ref{fig:1} shows (from left to right) G-band images of the FOV, map of velocities computed from local correlation tracking (LCT) techniques, and a divergence map for dataset G (top row) and dataset H (bottom row). This figure shows a spatial correspondence between the location of MBPs and lower velocities (red crosses in middle panels), and, in turn, with negative divergences (black crosses in right-most panels). Blue regions in the divergence maps trace intergranular lanes. This is in concordance with the presence of two populations of MBPs, with one group located in the network (with stronger magnetic fields and thus likely the larger MBPs showing lower velocities), and the other population of MBPs located in the internetwork (with weaker magnetic fields, harboring smaller MBPs which can move faster); however, a detailed analysis including magnetic field information is out of the scope of this analysis and will be covered in a future work.\\

The inspection of the results obtained by GREGOR and Hinode highlight the necessity for high spatial and temporal resolutions, as it is not possible to differentiate between the size of the two populations of MBPs with the Hinode instrument; however, it did become apparent with the GREGOR dataset. Moreover, at the same time the higher resolution of GREGOR helps to eliminate the misidentified population of small granules. Moreover, we can find higher velocities from the GREGOR telescope, which are related to the better temporal and spatial resolutions, as discussed in \citet{2010A&A...511A..39U}. The authors claim and verify in \citet{2012ASPC..454...55U} that a random walk motion observed under higher temporal cadence conditions would yield higher velocities as the length of the travelled path between two fixed time instances of the MBP increases. Such results are of particular value for MHD wave simulations as more rapid footpoint motions of the flux tube, i.e. MBP, will yield MHD waves carrying more energy into the higher atmosphere or possibly creating the recently observed magnetic field-line switchbacks \citep[e.g.][]{1993SoPh..143...49C,2009A&A...508..951V,2021ApJ...911...75M}.
Finally, with the hindsight of this work, it is even possible to find evidence for the two-component velocity distributions in earlier works such as \citet{2014SoPh..289.1543B}, in which the authors used the Dutch open telescope \citep[DOT;][]{2004A&A...413.1183R} data. Evidently, in their Fig.~2, depicting the magnitude of the MBP velocity, the second Rayleigh component has gone unnoticed by the authors, who, however, state that there is a widening discrepancy for the high velocity tail compared to their one-component Rayleigh fit. 

\section{Conclusions}
\label{sec:6}

In the present investigation, we studied the dynamics of MBPs on the solar photosphere for two quiet Sun regions, close to the central meridian of the solar disc. Observations from ground-based and space-borne solar telescopes were employed, specifically using the HiFI/GREGOR and the SOT/Hinode instruments. Unprecedented results for the size, diameter, and velocity of MBPs were found as a direct consequence of the excellent spatial resolution of the observations with the GREGOR telescope. The results of the characterisation of physical parameters of MBPs are in correspondence with others found in previous investigations  \citep[see][]{1994A&A...283..232M,2009A&A...498..289U,2010ApJ...722L.188C,2018ApJ...856...17L}.
The most important findings of the present investigation are related to the distribution of sizes, diameters, and velocities of MBPs. The results obtained with the two sets of observational data exhibit, for both area and diameter, a log-normal distribution, in agreement with previous observational and numerical studies \citep[e.g.][]{2010ApJ...722L.188C}. As suggested by \citet[][]{1988ApJ...327..451B}, the underlying fragmentation process might be responsible for log-normal distributions. In this work, the log-normal distributions of areas and diameters are found to be well described by two different components, each of which is proposed to correspond to a different population of solar features. As for velocities of MBPs, results show a Rayleigh distribution with the presence of two clear components, representing different groups of MBPs.
For dataset H (Hinode), a population of MBPs and small granules were found in the area and diameter results. This is due to difficulties arising from the lower spatial resolution of the telescope. On the other hand, in the velocity results two populations of MBPs were found. Similarly, for dataset G (GREGOR), two populations of MBPs were found for areas, diameters, and velocities, and no misidentifications (small granules) were found. This is by virtue of the prominent resolution of the telescope. Finally, another finding related to the effect of the intensity cut-off is that the cut-off percentage value of 80\% better describes the areas and diameters of the MBPs under study, which were not so large but consistent with those found in \citet{2009A&A...498..289U} and \citet{1980SoPh...65..207K}. In our case, using the commonly used definition cut-off at 50\% resulted in the misidentification of granules and excessively large values of MBPs for the area and diameter results.  

With the current study and the previous study of \citet[][]{2019MNRAS.488L..53K} describing two-component magnetic field strength distributions for MBPs, evidence is growing for a paradigm shift from a single population of MBPs to a more complex model. As a result, in the future we will be obliged to focus on two distinct types of MBPs, likely related to network and intranetwork magnetic fields and thus maybe even to the global magnetic field dynamo versus a shallow surface magnetic field dynamo. The new generation of solar telescopes, e.g. the Daniel K. Inouye Solar Telescope \citep[DKIST;][]{2021SoPh..296...70R} and the European Solar Telescope \citep[EST;][]{2019AdSpR..63.1389J} with larger apertures and, in turn, better spatial resolution, will shed more light on the two-component groups, enabling further confirmation of the characterisation of MBPs presented in this study.

\begin{acknowledgements}
      Part of this work was supported by Austrian FWF - Der Wissenschaftsfonds project number P27800 as well as the German
      Deut\-sche For\-schungs\-ge\-mein\-schaft, DFG\/ project
      number Ts~17/2--1. SVD acknowledges support by the Beyond Research Program between University of Graz and Universidad Nacional de Colombia. SJGM acknowledges the support of grants PGC2018-095832-B-I00 (MCIU) and ERC-2017- CoG771310-PI2FA (European Research Council). PG acknowledges the support of the project VEGA 2/0048/20. PZ acknowledges support from the Stefan Schwarz fund for postdoctoral researchers awarded by the Slovak Academy of Sciences. This research data leading to the results obtained has been supported by SOLARNET project that has received funding from the European Union’s Horizon 2020 research and innovation programme under grant agreement no 824135. The 1.5-meter GREGOR solar telescope was built by a German consortium under the leadership of the Kiepenheuer Institute for Solar Physics in Freiburg with the Leibniz Institute for Astrophysics Potsdam, the Institute for Astrophysics Göttingen, and the Max Planck Institute for Solar System Research in Göttingen as partners, and with contributions by the Instituto de Astrofísica de Canarias and the Astronomical Institute of the Academy of Sciences of the Czech Republic. The authors are grateful for the possibility to use Hinode data. Hinode is a Japanese mission developed and launched by ISAS/JAXA, collaborating with NAOJ as a domestic partner, NASA and UKSA as international partners. Scientific operation of the Hinode mission is conducted by the Hinode science team organized at ISAS/JAXA. This team mainly consists of scientists from institutes in the partner countries. Support for the post-launch operation is provided by JAXA and NAOJ (Japan), UKSA (U.K.), NASA, ESA, and NSC (Norway). We would like to express our gratitude to the anonymous referee for all the valuable comments that helped us you improve the presentation of the results and ideas highlighted in this work.

\end{acknowledgements} 

\bibliographystyle{aa}
\bibliography{Bibliography}

\end{document}